\documentclass[prb, reprint, superscriptaddress]{revtex4-2}
\usepackage{amsmath}
\usepackage{amssymb}
\usepackage{graphicx}
\usepackage[caption=false, position=top, singlelinecheck=off, justification=raggedright]{subfig}
\usepackage{multirow}
\usepackage{color}
\usepackage{pst-node}
\usepackage[unicode]{hyperref}
\hypersetup{
	unicode=true,         	
	colorlinks=true,      	
	linkcolor=blue,		   	
	citecolor=blue,       	
	urlcolor=blue		   	
}

\begin{document}

\title{Higgs mode mediated enhancement of interlayer transport in high-$T_c$ cuprate superconductors}

\author{Guido Homann}
\affiliation{Zentrum f\"ur Optische Quantentechnologien and Institut f\"ur Laserphysik, 
Universit\"at Hamburg, 22761 Hamburg, Germany}

\author{Jayson G. Cosme}
\affiliation{Zentrum f\"ur Optische Quantentechnologien and Institut f\"ur Laserphysik, 
Universit\"at Hamburg, 22761 Hamburg, Germany}
\affiliation{The Hamburg Centre for Ultrafast Imaging, Luruper Chaussee 149, 22761 Hamburg, Germany}
\affiliation{National Institute of Physics, University of the Philippines, Diliman, Quezon City 1101, Philippines}

\author{Junichi Okamoto}
\affiliation{Institute of Physics, University of Freiburg, Hermann-Herder-Strasse 3, 79104 Freiburg, Germany}
\affiliation{EUCOR Centre for Quantum Science and Quantum Computing, University of Freiburg, Hermann-Herder-Strasse 3, 79104 Freiburg, Germany}

\author{Ludwig Mathey}
\affiliation{Zentrum f\"ur Optische Quantentechnologien and Institut f\"ur Laserphysik, 
Universit\"at Hamburg, 22761 Hamburg, Germany}
\affiliation{The Hamburg Centre for Ultrafast Imaging, Luruper Chaussee 149, 22761 Hamburg, Germany}

\date{\today}
\begin{abstract}
We put forth a mechanism for enhancing the interlayer transport in cuprate superconductors, by optically driving plasmonic excitations along the $c$ axis with a frequency that is blue-detuned from the Higgs frequency. The plasmonic excitations induce a collective oscillation of the Higgs field which induces a parametric enhancement of the superconducting response, as we demonstrate with a minimal analytical model. Furthermore, we perform simulations of a particle-hole symmetric $U(1)$ lattice gauge theory and find good agreement with our analytical prediction. We map out the renormalization of the interlayer coupling as a function of the parameters of the optical field and demonstrate that the Higgs mode mediated enhancement can be larger than $50\%$.
\end{abstract}
\maketitle

\section{Introduction}
The observation of light-induced superconductivity in cuprates and organic salts has been associated with exciting lattice or molecular vibrations \cite{Fausti2011, Buzzi2020, Budden2020}. Related experiments on light-enhanced interlayer transport in the bilayer cuprate YBCO above and below the critical temperature $T_c$ have been reported in Refs.~\cite{Hu2014, Kaiser2014, Foerst2014, Cremin2019}. Several mechanisms for these observations have been proposed in Refs.~\cite{Mankowsky2014, Denny2015, Hoeppner2015, Raines2015, Patel2016, Okamoto2016, Okamoto2017}. These proposed mechanisms focus on inducing phononic motion and its influence on the superconducting response. Here, we propose to enhance the interlayer transport in cuprates by optically exciting Higgs oscillations. This collective motion of the Higgs field couples parametrically to the plasma field, which results in the enhancement of the superconducting response. Our primary example will be monolayer cuprates. We expect that similar results emerge for other lattice geometries as well. We demonstrate that the enhancement of the superconducting response, in particular the low-frequency behavior of the imaginary conductivity, is achieved via driving of the electric field along the $c$ axis with frequencies that are slightly blue-detuned from the Higgs frequency. Thus, we expand the scope of dynamical control of the superconducting state in the cuprates by exploiting nonlinear plasmonics \cite{Rajasekaran2016, Schlawin2017}.

In this paper, we first consider a two-mode model with a cubic coupling of the Higgs and plasma modes \cite{Homann2020}. Based on this minimal model, we provide an analytical expression for the Higgs mode mediated renormalization of the interlayer coupling in monolayer cuprates. We then extend our treatment to a $U(1)$ lattice gauge theory with inherent particle-hole symmetry and simulate the $c$-axis optical conductivity for different ratios of the Higgs and plasma frequencies at zero temperature. The numerical results confirm our analytical prediction, and we identify the optimal parameter regime for observing the Higgs mode mediated enhancement of interlayer transport. Finally, we discuss the feasibility of the effect and possible challenges brought by damping.

\begin{figure}[!b]
	\includegraphics[scale=1]{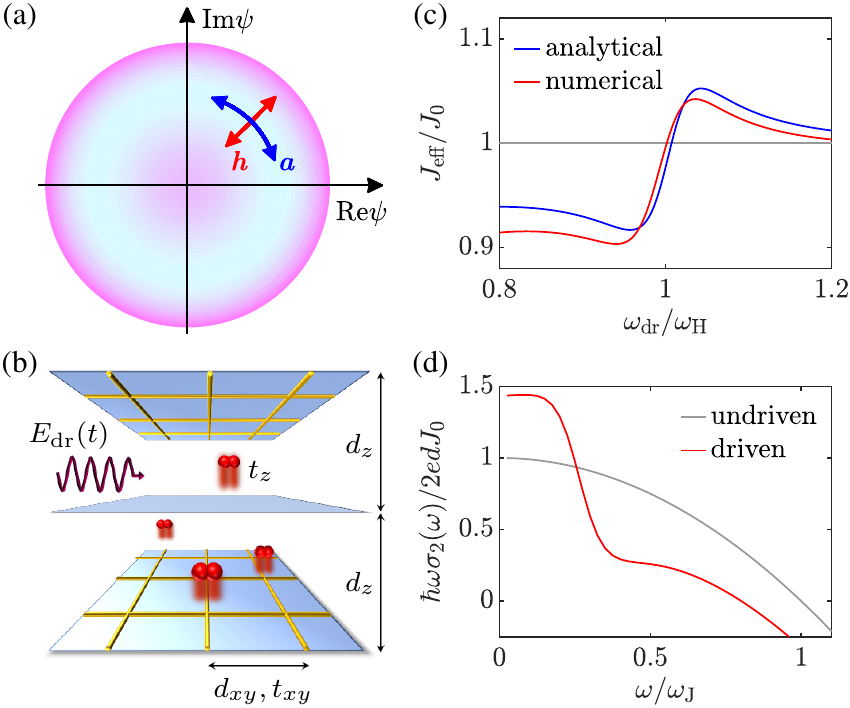}
	\caption{(a) Higgs and plasma modes of a monolayer cuprate superconductor, illustrated with a Mexican hat potential for the superconducting order parameter. (b) Schematic representation of a layered superconductor periodically driven by a $c$-axis polarized electric field with frequency $\omega_{\mathrm{dr}}$ and field strength $E_0$. (c) Effective interlayer coupling $J_{\mathrm{eff}}$ rescaled by its equilibrium value $J_0$. The field strength is fixed at $E_0= 100~\mathrm{kV/cm}$. (d) Numerical results for the imaginary conductivity $\sigma_2$. The driving parameters are $\omega_{\mathrm{dr}}= 1.05 \, \omega_{\mathrm{H}}$ and $E_0= 400~\mathrm{kV/cm}$. The cuprate considered in (c) and (d) has the Josephson plasma frequency $\omega_{\mathrm{J}}/2\pi=2~\mathrm{THz}$ and the Higgs frequency $\omega_{\mathrm{H}}/2\pi=6~\mathrm{THz}$.}
	\label{fig:fig1} 
\end{figure}

\section{Analytical prediction}
Expanding on previous works \cite{Koyama1996, Marel2001, Koyama2002, Harland2019}, we model a layered superconductor as a stack of intrinsic Josephson junctions. In addition to Josephson plasma resonances \cite{Dulic2001, Shibata1998, Basov2005}, recent experiments have revealed the existence of another fundamental excitation in cuprate superconductors, the Higgs mode \cite{Katsumi2018, Chu2020, Shimano2020}. This mode corresponds to amplitude oscillations of the superconducting order parameter $\psi$, which decouples from the plasma mode in a system with approximate particle-hole symmetry \cite{Varma2002, Pekker2015}. The two distinct low-energy modes of a monolayer cuprate superconductor are depicted in Fig.~\ref{fig:fig1}(a), where the phase of $\psi$ shall be interpreted as the gauge-invariant phase difference between adjacent layers. At zero momentum, the lowest-order coupling between the Higgs field $h$ and the unitless vector potential $a$ is given by the cubic interaction Lagrangian $\mathcal{L}_{\mathrm{int}} \sim a^2h$ \cite{Shimano2020, Sun2020}. The equations of motion corresponding to such a minimal model for describing the dynamics of a light-driven monolayer cuprate at zero temperature are
\begin{align}
	\ddot{a} + \gamma_{\mathrm{J}} \dot{a} + \omega_{\mathrm{J}}^2 a  + 2 \omega_{\mathrm{J}}^2 a h &= j, \label{eq:a1} \\
	\ddot{h} + \gamma_{\mathrm{H}} \dot{h} + \omega_{\mathrm{H}}^2 h + \alpha \omega_{\mathrm{J}}^2 a^2 &= 0, \label{eq:h1}
\end{align}
where $\omega_{\mathrm{H}}$ is the Higgs frequency, $\omega_{\mathrm{J}}$ is the plasma frequency, and $\gamma_{\mathrm{H}}$ and $\gamma_{\mathrm{J}}$ are damping coefficients. The capacitive coupling constant $\alpha$ is of the order of 1 in the cuprates \cite{Machida2004}. The interlayer current $j$ is induced by an external electric field. We use $\gamma_{\mathrm{H}}/2\pi= \gamma_{\mathrm{J}}/2\pi= 0.5~\mathrm{THz}$ and $\alpha=1$ unless stated otherwise \cite{Supp}, and we assume the $z$~axis to be aligned with the $c$~axis of the crystal.
 
For weak pump-probe strengths, we perform a perturbative expansion for $a$ and $h$ around their equilibrium values \cite{Supp}.
To calculate the $c$-axis optical conductivity $\sigma(\omega)$ in the driven state, we apply both driving and probing currents. We take the current as a first-order term in the expansion, i.e.,
\begin{equation}
j= \lambda \left(j_{\mathrm{dr},1} e^{-i \omega_{\mathrm{dr}} t} + j_{\mathrm{pr},1} e^{-i \omega_{\mathrm{pr}} t} + \mathrm{c.c.} \right) ,
\end{equation}
where $\lambda \ll 1$ is a small expansion parameter.
Hence, the leading contribution to $a$ is of first order and the leading contribution to $h$ is of second order. The coupling term $\sim ah$ gives a third order correction to $a$. Additionally, we assume that the probing frequency $\omega_{\mathrm{pr}}$ is much smaller than the driving frequency $\omega_{\mathrm{dr}}$ and the eigenfrequencies $\omega_{\mathrm{H}}$ and $\omega_{\mathrm{J}}$. Thus, we obtain an approximate expression for the Fourier component $a(\omega_{\mathrm{pr}})= \lambda a_{\mathrm{pr},1} + \lambda^3 a_{\mathrm{pr},3}$. The contributions are $a_{\mathrm{pr,1}} \approx j_{\mathrm{pr},1}/\omega_{\mathrm{J}}^2$ and
\begin{equation}
a_{\mathrm{pr},3} \approx \frac{ 4 \alpha |j_{\mathrm{dr},1}|^2 j_{\mathrm{pr},1} (\omega_{\mathrm{dr}}^2 - 3 \omega_{\mathrm{H}}^2 + i \gamma_{\mathrm{H}} \omega_{\mathrm{dr}}) }{ \omega_{\mathrm{H}}^2 (\omega_{\mathrm{dr}}^2 - \omega_{\mathrm{H}}^2 + i \gamma_{\mathrm{H}} \omega_{\mathrm{dr}}) \bigl[ (\omega_{\mathrm{dr}}^2 - \omega_{\mathrm{J}}^2)^2 + \gamma_{\mathrm{J}}^2 \omega_{\mathrm{dr}}^2 \bigr] } .
\end{equation}
This leads to the analytical prediction
\begin{widetext}
\begin{align}\label{eq:sigma}
	\omega_{\mathrm{pr}} \sigma(\omega_{\mathrm{pr}})= \frac{i \epsilon_z \epsilon_0 j(\omega_{\mathrm{pr}})}{a(\omega_{\mathrm{pr}})} &\approx \frac{ i \epsilon_z \epsilon_0 \omega_{\mathrm{J}}^2 \omega_{\mathrm{H}}^2 (\omega_{\mathrm{dr}}^2 - \omega_{\mathrm{H}}^2 + i \gamma_{\mathrm{H}} \omega_{\mathrm{dr}}) \bigl[ (\omega_{\mathrm{dr}}^2 - \omega_{\mathrm{J}}^2)^2 + \gamma_{\mathrm{J}}^2 \omega_{\mathrm{dr}}^2 \bigr] }{ \omega_{\mathrm{H}}^2 (\omega_{\mathrm{dr}}^2 - \omega_{\mathrm{H}}^2 + i \gamma_{\mathrm{H}} \omega_{\mathrm{dr}}) \bigl[ (\omega_{\mathrm{dr}}^2 - \omega_{\mathrm{J}}^2)^2 + \gamma_{\mathrm{J}}^2 \omega_{\mathrm{dr}}^2 \bigr] + 4 \alpha \omega_{\mathrm{J}}^2 |j_{\mathrm{dr}}|^2 (\omega_{\mathrm{dr}}^2 - 3 \omega_{\mathrm{H}}^2 + i \gamma_{\mathrm{H}} \omega_{\mathrm{dr}}) } .
\end{align}
\end{widetext}
where $\epsilon_z$ denotes the dielectric permittivity of the junctions, and $j_{\mathrm{dr}}= \lambda j_{\mathrm{dr},1}$ is the driving amplitude. 
We define an effective Josephson coupling \cite{Okamoto2016} based on the $1/\omega$ divergence of the conductivity:
\begin{align}\label{eq:jeff}
J_{\mathrm{eff}} = \frac{\hbar}{2ed_z}\mathrm{Im}[\omega_{\mathrm{pr}} \sigma(\omega_{\mathrm{pr}})]_{\omega_{\mathrm{pr}} \to 0},
\end{align}
with the interlayer spacing $d_z$. In the absence of driving, the Josephson coupling is $J_0 = \hbar \epsilon_z \epsilon_0 \omega^2_{\mathrm{J}}/(2ed_z)$ according to Eq.~\eqref{eq:sigma}. The analytical prediction for $J_{\mathrm{eff}}/J_0$ in the presence of driving is shown in Fig.~\ref{fig:fig1}(c). The key result of this work is the enhancement of the effective interlayer coupling when the pump frequency is slightly blue-detuned from the Higgs frequency. This enhancement phenomenon is due to parametric amplification. Indeed, Eq.~\eqref{eq:a1} takes the form of a parametric oscillator due to the two-wave mixing of drive and probe in Eq.~\eqref{eq:h1}, inducing amplitude oscillations at frequencies $2\omega_{\mathrm{dr}}$, $2\omega_{\mathrm{pr}}$, and $\omega_{\mathrm{dr}} \pm \omega_{\mathrm{pr}}$. The coupling of amplitude oscillations with $\omega_{\mathrm{dr}} \pm \omega_{\mathrm{pr}}$ to the drive amplifies the current response at the probing frequency. The numerical results in Figs.~\ref{fig:fig1}(c) and \ref{fig:fig1}(d), further highlighting the enhancement of interlayer transport, are obtained by simulating a full lattice gauge model discussed in the following.

\begin{figure*}[!t]
	\includegraphics[scale=1]{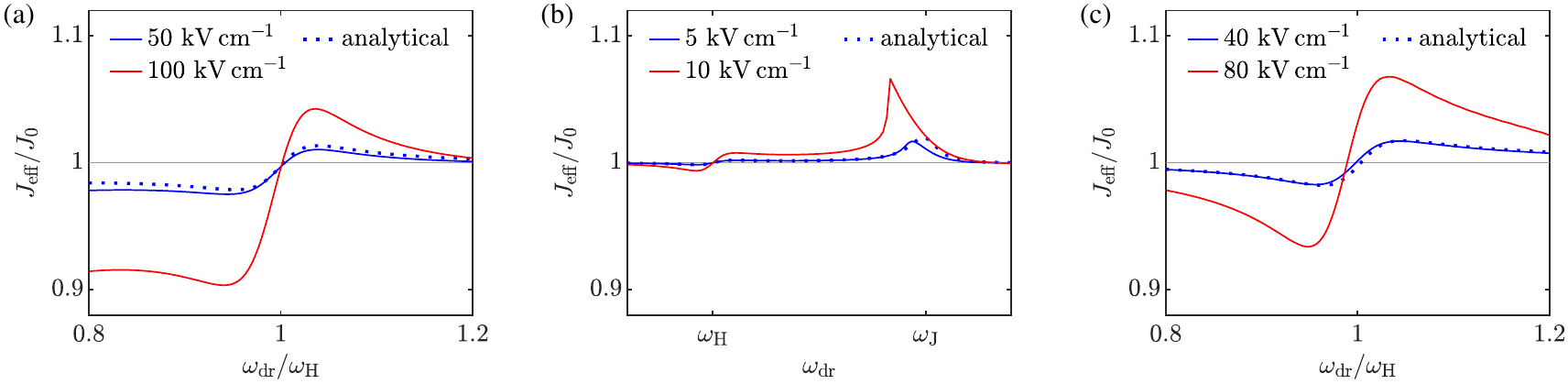}
	\caption{Effective interlayer coupling for three light-driven monolayer cuprate superconductors with different ratios $\omega_{\mathrm{J}}/\omega_{\mathrm{H}}$. (a) $\omega_{\mathrm{J}}<\omega_{\mathrm{H}}$,  (b) $\omega_{\mathrm{H}}<\omega_{\mathrm{J}} <\sqrt{3}\omega_{\mathrm{H}}$, (c) $\sqrt{3}\omega_{\mathrm{H}}<\omega_{\mathrm{J}} $. Each panel shows the dependence on the driving frequency $\omega_{\mathrm{dr}}$ for two fixed values of the field strength $E_0$. Solid lines correspond to numerical results, and dotted lines indicate analytical results for the lower field strength. In all cases, the Higgs frequency is fixed at $\omega_{\mathrm{H}}/2\pi=6~\mathrm{THz}$, while the plasma frequency $\omega_{\mathrm{J}}/2\pi$ is varied: (a) $2~\mathrm{THz}$, (b) $9~\mathrm{THz}$, and (c) $15~\mathrm{THz}$.}
	\label{fig:fig2} 
\end{figure*}

\section{Lattice gauge model}
We now turn to our relativistic $U(1)$ lattice gauge theory in three dimensions, which is inherently particle-hole symmetric. The layered structure of cuprate superconductors is encoded in the lattice parameters. Our approach allows us to explicitly simulate the coupled dynamics of the order parameter of the superconducting state $\psi_{\mathbf{r}}(t)$ and the electromagnetic field $\mathbf{A}_{\mathbf{r}}(t)$ at temperatures below $T_c$. To this end, we describe the Cooper pairs as a condensate of interacting bosons with charge $-2e$, represented by the complex field $\psi_{\mathbf{r}}(t)$. The time-independent part of our model Lagrangian has the form of the Ginzburg-Landau free energy \cite{Ginzburg1950}, discretized on an anisotropic lattice.
We model the layered structure of high-$T_c$ cuprates using an anisotropic lattice geometry as illustrated in Fig.~\ref{fig:fig1}(b).
The in-plane discretization length $d_{xy}$ constitutes a short-range cutoff around the coherence length, while the interlayer spacing $d_z$ is the distance between the CuO$_2$ planes in the crystal.
Each component of the vector potential $A_{s,\mathbf{r}}(t)$ is located at half a lattice site from site $\mathbf{r}$ in the $s$ direction, where $s \in \{x,y,z\}$. According to the Peierls substitution, it describes the averaged electric field along the bond of a plaquette in Fig.~\ref{fig:fig1}(b).

We discretize space by mapping it on a lattice and implement the compact $U(1)$ lattice gauge theory in the time continuum limit \cite{Kogut1979}. The Lagrangian of the lattice gauge model is
\begin{equation} \label{eq:Lagrangian}
	\mathcal{L} = \mathcal{L}_{\mathrm{sc}} + \mathcal{L}_{\mathrm{em}} + \mathcal{L}_{\mathrm{kin}} .
\end{equation}
The first term is the $|\psi|^4$ model of the superconducting condensate in the absence of Cooper pair tunneling:
\begin{equation}\label{eq:LSC}
	\mathcal{L}_{\mathrm{sc}} = \sum_{\mathbf{r}} K \hbar^2 | \partial_{t} \psi_{\mathbf{r}} |^2 + \mu | \psi_{\mathbf{r}} |^2 - \frac{g}{2} | \psi_{\mathbf{r}} |^4 ,
\end{equation}
where $\mu$ is the chemical potential, and $g$ is the interaction strength. This Lagrangian is particle-hole symmetric due to its invariance under $\psi_{\mathbf{r}} \rightarrow \psi_{\mathbf{r}}^*$. 
The coefficient $K$ describes the magnitude of the dynamical term \cite{Pekker2015,Shimano2020}.

The electromagnetic part $\mathcal{L}_{\mathrm{em}}$ is the discretized form of the free-field Lagrangian:
\begin{equation} \label{eq:LEM}
\mathcal{L}_{\mathrm{em}} = 
\sum_{s,\mathbf{r}} \frac{\epsilon_s \epsilon_0}{2} E_{s,\mathbf{r}}^2 - \frac{1}{\mu_0 \beta_s^2} \Bigl[1 - \cos\bigl(\beta_s B_{s,\mathbf{r}} \bigr) \Bigr] .
\end{equation}
Here $E_{s,\mathbf{r}}$ denotes the $s$~component of the electric field, and $\epsilon_s$ is the dielectric permittivity along that axis. The magnetic field components $B_{s,\mathbf{r}}$ follow from the finite-difference representation of $\mathbf{\nabla} \times \mathbf{A}$. The temporal and spatial arrangement of the electromagnetic field is consistent with the finite-difference time-domain (FDTD) method for solving Maxwell's equations \cite{Yee1966}. Note that we choose the temporal gauge for our calculations, i.e., $E_{s,\mathbf{r}} = -\partial_{t} A_{s,\mathbf{r}}$. The coefficients for the magnetic field are $\beta_{x}= \beta_{y}= 2ed_{xy}d_z/\hbar$ and $\beta_z= 2ed_{xy}^2/\hbar$.

The nonlinear coupling between the Higgs field and the electromagnetic field derives from the tunneling term 
\begin{equation} \label{eq:LEK}
\mathcal{L}_{\mathrm{kin}} = - \sum_{s,\mathbf{r}} t_s |\psi_{\mathbf{r'}(s)} - \psi_{\mathbf{r}} e^{i a_{s,\mathbf{r}}}|^2 ,
\end{equation}
where $\mathbf{r}'(s)$ denotes the neighboring lattice site of $\mathbf{r}$ in the positive $s$~direction. The unitless vector potential $a_{s,\mathbf{r}}= -2e d_s A_{s,\mathbf{r}}/\hbar$ couples to the phase of the superconducting field, ensuring the local gauge-invariance of $\mathcal{L}_{\mathrm{kin}}$. The in-plane tunneling coefficient is $t_{xy}$, and the interlayer tunneling coefficient is $t_z$.

We numerically solve the equations of motion derived from the Lagrangian, including damping terms. We employ periodic boundary conditions and integrate the differential equations using Heun's method with a step size $\Delta t= 2.5~\mathrm{as}$.
Here, we focus on zero temperature, where the in-plane dynamics is silent. An example of Higgs mode mediated enhancement at nonzero temperature is included in the Supplemental Material \cite{Supp}.

We drive the system by adding $(\omega_{\mathrm{dr}} E_0/\epsilon_z) \mathrm{sin}(\omega_{\mathrm{dr}} t)$ to the equations of motion for the vector potential $A_{z,\mathbf{r}}(t)$ on all interlayer bonds, which describes a spatially homogeneous driving field. 
Note that Eqs.~\eqref{eq:a1} and \eqref{eq:h1} can be derived as the Euler-Lagrange equations of the Lagrangian~\eqref{eq:Lagrangian} at zero temperature. In that case, the  fields are uniform in the bulk, i.e., $\psi_{\mathbf{r}} \equiv \psi$, $A_{x,\mathbf{r}} = A_{y,\mathbf{r}} \equiv 0$, and $A_{z,\mathbf{r}} \equiv A$. The equations of motion read
\begin{equation}
\partial_{t}^2 A= \frac{2 e d_z t_z}{i \hbar \epsilon_z \epsilon_0} |\psi|^2 \left(e^{ia} - e^{-ia} \right) - \gamma_{\mathrm{J}} \partial_{t} A + \frac{\omega_{\mathrm{dr}} E_0}{\epsilon_z} \mathrm{sin}(\omega_{\mathrm{dr}} t)
\end{equation}
and
\begin{equation}
\partial_{t}^2 \psi= \frac{\mu - g |\psi|^2 + t_z \left( e^{ia} + e^{-ia} -2 \right)}{K\hbar^2} \psi  -\gamma_{\mathrm{H}} \partial_{t} \psi ,
\end{equation}
where $a= -2ed_zA/\hbar$. To recover Eqs.~\eqref{eq:a1} and \eqref{eq:h1}, the order parameter is expanded around its equilibrium value $\psi_0= \sqrt{\mu/g}$, i.e., $\psi= \psi_0 + h$, and only linear terms in $a$ and $h$ except for the coupling term $\sim ah$ are retained. Thus, one can identify the plasma frequency with $\omega_{\mathrm{J}}= \sqrt{t_z/\alpha K \hbar^2}$ and the Higgs frequency with $\omega_{\mathrm{H}}= \sqrt{2 \mu/K \hbar^2}$, where $\alpha= (\epsilon_z \epsilon_0)/(8 K \psi_0^2 e^2 d_z^2)$ is the capacitive coupling constant of the $c$-axis junctions. The drive induces a current with Fourier amplitude $|j_{\mathrm{dr}}|= e d_z \omega_{\mathrm{dr}} E_0/\hbar \epsilon_z$.

\section{Numerical results}
In the following, we present our numerical results. We evaluate the effective interlayer coupling based on the optical conductivity at $\omega_{\mathrm{pr}}= \omega_{\mathrm{H}}/120$.
For weak driving, we find decent agreement between the analytical prediction in Eq.~\eqref{eq:sigma} and the numerical results of the full lattice gauge model, as shown in Fig.~\ref{fig:fig2}. The deviations are due to higher-order terms ignored in the minimal model and the perturbative expansion. They grow with increasing field strength. Nevertheless, our simulations demonstrate that the enhancement effect persists for strong driving, even in the presence of higher-order nonlinearities, fully included in our $U(1)$ lattice gauge theory.

\begin{figure}[!t]
	\includegraphics[scale=1]{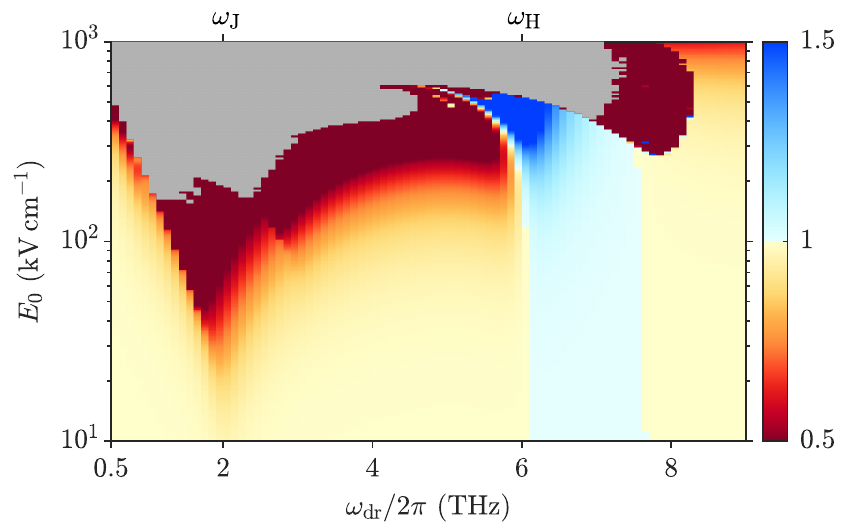}
	\caption{Dependence of the effective interlayer coupling $J_{\mathrm{eff}}/J_0$ on the driving frequency $\omega_{\mathrm{dr}}$ and the field strength $E_0$. The gray area marks the heating regime.}
	\label{fig:fig3} 
\end{figure}

We find that the renormalization of the interlayer coupling does not only depend on the driving parameters, but also on the ratio of the Higgs frequency and the plasma frequency of the system.
Our main proposal consists of driving the superconductor slightly blue-detuned from the Higgs frequency $\omega_{\mathrm{H}}$. This mechanism is effective for all ratios of $\omega_{\mathrm{J}}/\omega_{\mathrm{H}}$. As we discuss below, there is a second regime in which dynamical stabilization can be achieved, if the system fulfills the requirement $\omega_{\mathrm{H}} < \omega_{\mathrm{J}} < \sqrt{3}\omega_{\mathrm{H}}$.

Figure~\ref{fig:fig3} displays the renormalized interlayer coupling as a function of the driving parameters for a monolayer cuprate with $\omega_{\mathrm{J}} < \omega_{\mathrm{H}}$ [the same system as in Figs.~\ref{fig:fig1} and \ref{fig:fig2}(a)]. Consistent with our analytical prediction, the interlayer transport is enhanced for driving frequencies blue-detuned from the Higgs frequency, while it is diminished on the red-detuned side, as immediately apparent for weak driving.
In general, higher field strengths amplify the suppression/enhancement effects and additionally renormalize the Higgs frequency and the plasma frequency. The frequency renormalization of the Higgs mode results in the bending of the enhancement regime towards lower driving frequencies for larger driving fields. This observation reflects the general behavior of nonlinear oscillators to display amplitude-dependent eigenfrequencies \cite{Landau1976}.
We emphasize that the interlayer coupling can be increased by more than $50\%$ in this example. The strongest suppression of interlayer transport occurs for driving close to the plasma frequency. This is generally the case if $\omega_{\mathrm{J}} < \omega_{\mathrm{H}}$ or $\omega_{\mathrm{J}} > \sqrt{3} \omega_{\mathrm{H}}$.

The enhancement and suppression effects are limited by heating that dominates for larger field strengths (see also Ref.~\cite{Homann2020}). We identify the heating regime based on the condition that the condensate is completely depleted. Specifically, we observe the driven dynamics for 100 ps and apply the condition $\mathrm{min}(|\psi|/|\psi_0|) < 10^{-3}$ to determine unstable states. We note that the heating regime has a similar shape as the parameter set for which no stable solutions can be found by applying the harmonic balance method with ten harmonics \cite{Supp, Slater2017}.

Within our analytical solution for the optical conductivity in Eq.~\eqref{eq:sigma}, we have determined an upper boundary for the driving frequency of $\sqrt{3} \omega_{\mathrm{H}}$ for the enhancement effect to occur. At this boundary, the second term in the denominator of Eq.~\eqref{eq:sigma} switches sign. Our simulations confirm this prediction for $\omega_{\mathrm{H}} < \omega_{\mathrm{J}}$. However, as visible in Fig.~\ref{fig:fig3}, an additional suppression lowers this upper bound for superconductors with $\omega_{\mathrm{J}} < (\sqrt{3} -1) \omega_{\mathrm{H}}$. Here, the enhancement regime is approximately limited by the resonance frequency of the time crystalline state at $\omega_{\mathrm{dr}} = \omega_{\mathrm{H}} + \omega_{\mathrm{J}}$ \cite{Homann2020}. This modified upper bound derives from higher-order terms not included in the analytical solution.

\begin{figure}[!b]
	\includegraphics[scale=1]{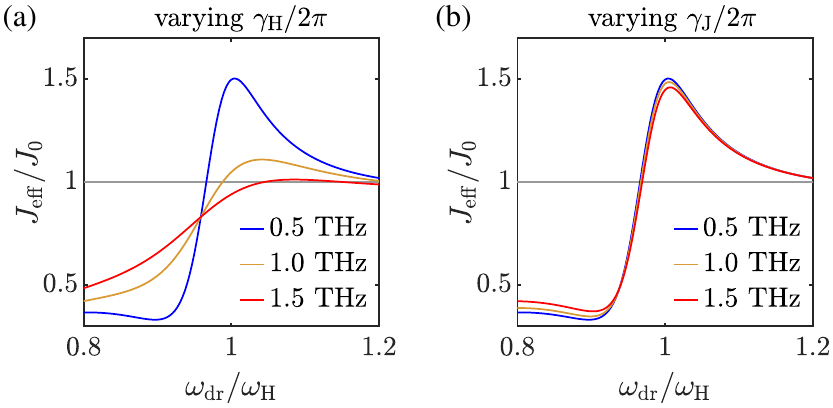}
	\caption{Dependence of the Higgs mode mediated renormalization of the interlayer coupling on the damping coefficients. (a) $\gamma_{\mathrm{H}}$ is varied while $\gamma_{\mathrm{J}}/2\pi= 0.5~\mathrm{THz}$ is fixed. (b) $\gamma_{\mathrm{J}}$ is varied while $\gamma_{\mathrm{H}}/2\pi= 0.5~\mathrm{THz}$ is fixed. The cuprate with Josephson plasma frequency $\omega_{\mathrm{J}}/2\pi=2~\mathrm{THz}$ and Higgs frequency $\omega_{\mathrm{H}}/2\pi=6~\mathrm{THz}$ is driven with the field strength $E_0= 300~\mathrm{kV/cm}$.}
	\label{fig:fig4} 
\end{figure}

We continue our analysis by varying the damping coefficients, as shown in Fig.~\ref{fig:fig4}. Studying higher values of $\gamma_{\mathrm{H}}$ is particularly interesting because the damping of the Higgs mode is typically strong in cuprate superconductors \cite{Katsumi2018, Chu2020}. It can be seen in Fig.~\ref{fig:fig4}(a) that increasing $\gamma_{\mathrm{H}}$ significantly decreases the enhancement of the interlayer coupling for a given field strength. By contrast, stronger damping of the plasma mode has a negligible effect. In the Supplemental Material \cite{Supp}, we provide a parameter scan of the renormalized interlayer coupling with higher damping coefficients $\gamma_{\mathrm{H}}/2\pi= \gamma_{\mathrm{J}}/2\pi= 1~\mathrm{THz}$. Compared to Fig.~\ref{fig:fig3}, we find that the parameter regime with an enhancement of more than 10\% is smaller and shifted to higher field strengths.

Finally, we consider cuprates with $\omega_{\mathrm{H}} < \omega_{\mathrm{J}} < \sqrt{3}\omega_{\mathrm{H}}$. In this case, the previous suppression of interlayer transport for $\omega_{\mathrm{dr}} \approx \omega_{\mathrm{J}}$ switches to strong enhancement, as exemplified in Fig.~\ref{fig:fig2}(b). Therefore, we propose to drive these particular systems near the plasma frequency $\omega_{\mathrm{J}}$.
In typical monolayer cuprates, such as LSCO, the superconducting gap $2\Delta$ is larger than the Josephson plasma energy $\hbar \omega_{\mathrm{J}}$ \cite{Dordevic2003, Deutscher1999, Hashimoto2007}. At low temperatures, the Higgs frequency approximately equals $2 \Delta/\hbar$ \cite{Shimano2020, Yang2020}, so it is larger than the Josephson plasma frequency in these materials, i.e., $\omega_{\mathrm{J}} < \omega_{\mathrm{H}}$. However, while the temperature dependence of the Higgs mode in cuprate superconductors is the subject of debate \cite{Katsumi2018, Chu2020, Puviani2020, Gabriele2021}, the case $\omega_{\mathrm{H}} < \omega_{\mathrm{J}} < \sqrt{3}\omega_{\mathrm{H}}$ might be realized for higher temperatures. For these temperatures, stronger damping and thermal fluctuations might suppress or reduce the enhancement mechanism. This regime will be discussed elsewhere. Further decay channels of the Higgs mode have been studied in Refs.~\cite{Peronaci2015, Murakami2016}.

\section{Conclusion}
In conclusion, we propose a mechanism for light-enhanced interlayer transport in cuprate superconductors by optically exciting Higgs oscillations which then induce a parametric amplification of the superconducting response. Both our analytical and numerical calculations show that the superconducting response of a monolayer cuprate is significantly amplified when the optical driving is slightly blue-detuned from the Higgs frequency. Our calculations demonstrate that the regime of driving parameters, for which a significant Higgs mode mediated enhancement of interlayer transport is achieved, crucially depends on the damping of the Higgs mode. Therefore, we propose to verify this effect first for low temperatures. The enhancement mechanism presented in this work is broadly applicable to cuprate superconductors because it does not rely on the existence of suitable phonons. Instead, the light-driven renormalization of interlayer transport is mediated by Higgs oscillations of the Cooper pair condensate itself. This effect amounts to dynamical control of a functionality in high-temperature superconductors, utilizing the intrinsic collective modes of these materials and their nonlinear coupling.

\vspace{0.5cm}
\begin{acknowledgments}
This work is supported by the Deutsche Forschungsgemeinschaft (DFG) in the framework of SFB~925, Project No.~170620586, and the Cluster of Excellence ``Advanced Imaging of Matter" (EXC~2056), Project No.~390715994. J.O. acknowledges support by the Georg H. Endress Foundation.
\end{acknowledgments}

\appendix*

\section{Model parameters}
Table~\ref{tab:parameters} summarizes the parameters of the monolayer cuprate superconductors studied in this paper. Our parameter choice of $\mu$ and $g$ implies an equilibrium condensate density $n_0= \mu/g = 2 \times 10^{21}~\mathrm{cm^{-3}}$ at zero temperature. The capacitive coupling constant is given by
\begin{equation}
	\alpha= \frac{g \epsilon_z \epsilon_0}{8 \mu K e^2 d_z^2} = 1 .
\end{equation}
For the $c$-axis plasma frequency, we consider the three cases $\omega_{\mathrm{J}}/2\pi= 2~\mathrm{THz}$, $\omega_{\mathrm{J}}/2\pi= 9~\mathrm{THz}$, and $\omega_{\mathrm{J}}/2\pi= 15~\mathrm{THz}$. The Higgs frequency is $\omega_{\mathrm{H}}/2\pi= 6~\mathrm{THz}$ in each case.
\begin{table}[!b]
	\caption{Model parameters used in the simulations.}
	\renewcommand{\arraystretch}{1.5}
	\begin{tabular}{lr}
		\hline
		$K~(\text{meV}^{-1})$ & $1.38 \times 10^{-5}$ \\
		$\mu~(\text{meV})$ & $4.24 \times 10^{-3}$ \\
		$g~(\text{meV} \, \text{\AA}^3)$ & 2.12 \\
		\hline
		$\epsilon_{xy}$ & 1 \\
		$\epsilon_z$ & 4 \\
		$d_{xy}~(\text{\AA})$ & 20 \\
		$d_z~(\text{\AA})$ & 10 \\
		$t_{xy}~(\text{meV})$ & $2.2 \times 10^{-1}$ \\
		$t_z~(\text{meV})$ & $\qquad 9.44 \times 10^{-4}$, $1.91 \times 10^{-2}$, $5.31 \times 10^{-2}$ \\
		\hline
	\end{tabular}
	\renewcommand{\arraystretch}{1}
	\label{tab:parameters}
\end{table}

\bibliography{biblio}

\end{document}


\title{Supplemental Material for\\Higgs mode mediated enhancement of interlayer transport in high-$T_c$ cuprate superconductors}

\author{Guido Homann}
\affiliation{Zentrum f\"ur Optische Quantentechnologien and Institut f\"ur Laserphysik, 
	Universit\"at Hamburg, 22761 Hamburg, Germany}

\author{Jayson G. Cosme}
\affiliation{Zentrum f\"ur Optische Quantentechnologien and Institut f\"ur Laserphysik, 
	Universit\"at Hamburg, 22761 Hamburg, Germany}
\affiliation{The Hamburg Centre for Ultrafast Imaging, Luruper Chaussee 149, 22761 Hamburg, Germany}
\affiliation{National Institute of Physics, University of the Philippines, Diliman, Quezon City 1101, Philippines}

\author{Junichi Okamoto}
\affiliation{Institute of Physics, University of Freiburg, Hermann-Herder-Strasse 3, 79104 Freiburg, Germany}
\affiliation{EUCOR Centre for Quantum Science and Quantum Computing, University of Freiburg, Hermann-Herder-Strasse 3, 79104 Freiburg, Germany}

\author{Ludwig Mathey}
\affiliation{Zentrum f\"ur Optische Quantentechnologien and Institut f\"ur Laserphysik, 
	Universit\"at Hamburg, 22761 Hamburg, Germany}
\affiliation{The Hamburg Centre for Ultrafast Imaging, Luruper Chaussee 149, 22761 Hamburg, Germany}

\maketitle
\tableofcontents

\clearpage
\section{Analytical derivation of the Higgs mode mediated enhancement of interlayer transport}
\label{sec:analytical}
Here, we consider a monolayer cuprate superconductor at zero temperature. Since the system is driven along the $c$~axis, it exhibits no in-plane dynamics. We rewrite Eqs.~(11) and (12) from the main text as
\begin{align}
	\ddot{a} + \gamma_{\mathrm{J}} \dot{a} + \omega_{\mathrm{J}}^2 \sin(a)(1+h)^2 &= j , \label{eq:eom1_full} \\
	\ddot{h} + \gamma_{\mathrm{H}} \dot{h} + \omega_{\mathrm{H}}^2 \left( h + \frac{3}{2}h^2 + \frac{1}{2} h^3 \right)+ 2\alpha \omega_{\mathrm{J}}^2\left[1-\cos(a) \right](1+h) &= 0 . \label{eq:eom2_full}
\end{align}
Neglecting all nonlinear terms except for the quadratic coupling between the Higgs field $h$ and the unitless vector potential $a$, we find
\begin{align}
	\ddot{a} + \gamma_{\mathrm{J}} \dot{a} + \omega_{\mathrm{J}}^2 a + 2 \omega_{\mathrm{J}}^2 a h &= j , \label{eq:eom1_quad} \\
	\ddot{h} + \gamma_{\mathrm{H}} \dot{h} + \omega_{\mathrm{H}}^2 h + \alpha \omega_{\mathrm{J}}^2 a^2 &= 0 . \label{eq:eom2_quad}
\end{align}
Now, we expand $j$, $a$, and $h$ in the form
\begin{equation}
f= f^{(0)} + \lambda f^{(1)} + \lambda^2 f^{(2)} + \lambda^3 f^{(3)} + \mathcal{O}(\lambda^4) ,
\end{equation}
where $\lambda \ll 1$ is a small expansion parameter. We take the current $j$ induced by driving and probing as
\begin{equation}
j^{(1)}= j_{\mathrm{dr},1} e^{- i \omega_{\mathrm{dr}} t} + j_{\mathrm{pr},1} e^{- i \omega_{\mathrm{pr}} t} + \mathrm{c.c.} .
\end{equation}
Hence, there are no zeroth order contributions and we obtain
\begin{align}
a^{(1)} &= a_{\mathrm{dr},1} e^{- i \omega_{\mathrm{dr}} t} + a_{\mathrm{pr},1} e^{- i \omega_{\mathrm{pr}} t} + \mathrm{c.c.} , \\
h^{(1)} &= 0
\end{align}
in first order, where
\begin{align}
a_{\mathrm{dr,1}} &= \frac{j_{\mathrm{dr},1}}{\omega_{\mathrm{J}}^2 - \omega_{\mathrm{dr}}^2 - i \gamma_{\mathrm{J}} \omega_{\mathrm{dr}}} , \\
a_{\mathrm{pr,1}} &= \frac{j_{\mathrm{pr},1}}{\omega_{\mathrm{J}}^2 - \omega_{\mathrm{pr}}^2 - i \gamma_{\mathrm{J}} \omega_{\mathrm{pr}}} . \label{eq:den1}
\end{align}
In second order, we have
\begin{align}
a^{(2)} &= 0 , \\
h^{(2)} &= h_0 + h_1 e^{- 2 i \omega_{\mathrm{dr}} t} + h_2 e^{- 2 i \omega_{\mathrm{pr}} t} + h_3 e^{- i(\omega_{\mathrm{dr}} - \omega_{\mathrm{pr}}) t} + h_4 e^{- i(\omega_{\mathrm{dr}} + \omega_{\mathrm{pr}}) t} + \mathrm{c.c.} ,
\end{align}
where
\begin{align}
h_0 &= - \frac{2 \alpha \omega_{\mathrm{J}}^2}{\omega_{\mathrm{H}}^2} \big( |a_{\mathrm{dr},1}|^2 + |a_{\mathrm{pr},1}|^2 \big) , \\
h_1 &= \frac{\alpha \omega_{\mathrm{J}}^2 a_{\mathrm{dr},1}^2}{4 \omega_{\mathrm{dr}}^2 - \omega_{\mathrm{H}}^2 + 2 i \gamma_{\mathrm{H}} \omega_{\mathrm{dr}}} , \\
h_2 &= \frac{\alpha \omega_{\mathrm{J}}^2 a_{\mathrm{pr},1}^2}{4 \omega_{\mathrm{pr}}^2 - \omega_{\mathrm{H}}^2 + 2 i \gamma_{\mathrm{H}} \omega_{\mathrm{pr}}} , \\
h_3 &= \frac{2 \alpha \omega_{\mathrm{J}}^2 a_{\mathrm{dr},1} a_{\mathrm{pr},1}^*}{(\omega_{\mathrm{dr}} - \omega_{\mathrm{pr}})^2 - \omega_{\mathrm{H}}^2 + i \gamma_{\mathrm{H}} (\omega_{\mathrm{dr}} - \omega_{\mathrm{pr}})} , \label{eq:den2} \\
h_4 &= \frac{2 \alpha \omega_{\mathrm{J}}^2 a_{\mathrm{dr},1} a_{\mathrm{pr},1}}{(\omega_{\mathrm{dr}} + \omega_{\mathrm{pr}})^2 - \omega_{\mathrm{H}}^2 + i \gamma_{\mathrm{H}} (\omega_{\mathrm{dr}} + \omega_{\mathrm{pr}})} . \label{eq:den3}
\end{align}
In third order, we find the following correction for the vector potential at the probing frequency:
\begin{equation} \label{eq:den4}
a_{\mathrm{pr},3}=  \frac{2 \omega_{\mathrm{J}}^2 \big( h_0 a_{\mathrm{pr},1} + h_2 a_{\mathrm{pr},1}^* + h_3^* a_{\mathrm{dr},1} + h_4 a_{\mathrm{dr},1}^* \big)}{\omega_{\mathrm{pr}}^2 - \omega_{\mathrm{J}}^2 + i \gamma_{\mathrm{J}} \omega_{\mathrm{pr}}} .
\end{equation}
Using $|j_{\mathrm{pr},1}| \ll |j_{\mathrm{dr},1}|$ and $\omega_{\mathrm{pr}} \ll \omega_{\mathrm{dr}}, \omega_{\mathrm{H}}, \omega_{\mathrm{J}}$, we can neglect terms proportional to $|a_{\mathrm{pr},1}|^2 a_{\mathrm{pr},1}$ and simplify the denominators of Eqs.~\eqref{eq:den1}, \eqref{eq:den2}, \eqref{eq:den3}, and \eqref{eq:den4}:
\begin{align}
\begin{split}
a_{\mathrm{pr},3} &\approx  \frac{4 \alpha \omega_{\mathrm{J}}^2 |a_{\mathrm{dr},1}|^2 a_{\mathrm{pr},1}}{\omega_{\mathrm{H}}^2} \Biggl( 1 - \frac{2 \omega_{\mathrm{H}}^2}{\omega_{\mathrm{dr}}^2 - \omega_{\mathrm{H}}^2 + i \gamma_{\mathrm{H}} \omega_{\mathrm{dr}}} \Biggr) \\
&\approx \frac{ 4 \alpha |j_{\mathrm{dr},1}|^2 j_{\mathrm{pr},1} (\omega_{\mathrm{dr}}^2 - 3 \omega_{\mathrm{H}}^2 + i \gamma_{\mathrm{H}} \omega_{\mathrm{dr}}) }{ \omega_{\mathrm{H}}^2 (\omega_{\mathrm{dr}}^2 - \omega_{\mathrm{H}}^2 + i \gamma_{\mathrm{H}} \omega_{\mathrm{dr}}) \bigl[ (\omega_{\mathrm{dr}}^2 - \omega_{\mathrm{J}}^2)^2 + \gamma_{\mathrm{J}}^2 \omega_{\mathrm{dr}}^2 \bigr] } .
\end{split}
\end{align}
Thus, we obtain
\begin{align}
\begin{split}
\omega_{\mathrm{pr}} \sigma(\omega_{\mathrm{pr}}) &= \frac{i \epsilon_z \epsilon_0 j (\omega_{\mathrm{pr}})}{a (\omega_{\mathrm{pr}})} \\
&= \frac{i \epsilon_z \epsilon_0 \lambda j_{\mathrm{pr},1}}{\lambda a_{\mathrm{pr},1} + \lambda^3 a_{\mathrm{pr},3}} \\
&\approx \frac{ i \epsilon_z \epsilon_0 \omega_{\mathrm{J}}^2 \omega_{\mathrm{H}}^2 (\omega_{\mathrm{dr}}^2 - \omega_{\mathrm{H}}^2 + i \gamma_{\mathrm{H}} \omega_{\mathrm{dr}}) \bigl[ (\omega_{\mathrm{dr}}^2 - \omega_{\mathrm{J}}^2)^2 + \gamma_{\mathrm{J}}^2 \omega_{\mathrm{dr}}^2 \bigr] }{ \omega_{\mathrm{H}}^2 (\omega_{\mathrm{dr}}^2 - \omega_{\mathrm{H}}^2 + i \gamma_{\mathrm{H}} \omega_{\mathrm{dr}}) \bigl[ (\omega_{\mathrm{dr}}^2 - \omega_{\mathrm{J}}^2)^2 + \gamma_{\mathrm{J}}^2 \omega_{\mathrm{dr}}^2 \bigr] + 4 \alpha \omega_{\mathrm{J}}^2 |j_{\mathrm{dr}}|^2 (\omega_{\mathrm{dr}}^2 - 3 \omega_{\mathrm{H}}^2 + i \gamma_{\mathrm{H}} \omega_{\mathrm{dr}}) } ,
\end{split}
\end{align}
with the original driving amplitude $j_{\mathrm{dr}}= \lambda j_{\mathrm{dr},1}$. Taking $j_{\mathrm{dr}}=0$ leads to the equilibrium solution
\begin{equation}
\omega_{\mathrm{pr}} \sigma(\omega_{\mathrm{pr}}) \approx i \epsilon_z \epsilon_0 \omega_{\mathrm{J}}^2 .
\end{equation}

\section{Lattice gauge model}
\label{sec:parameters}
\begin{figure}[!h]
	\centering
	\includegraphics[scale=0.83]{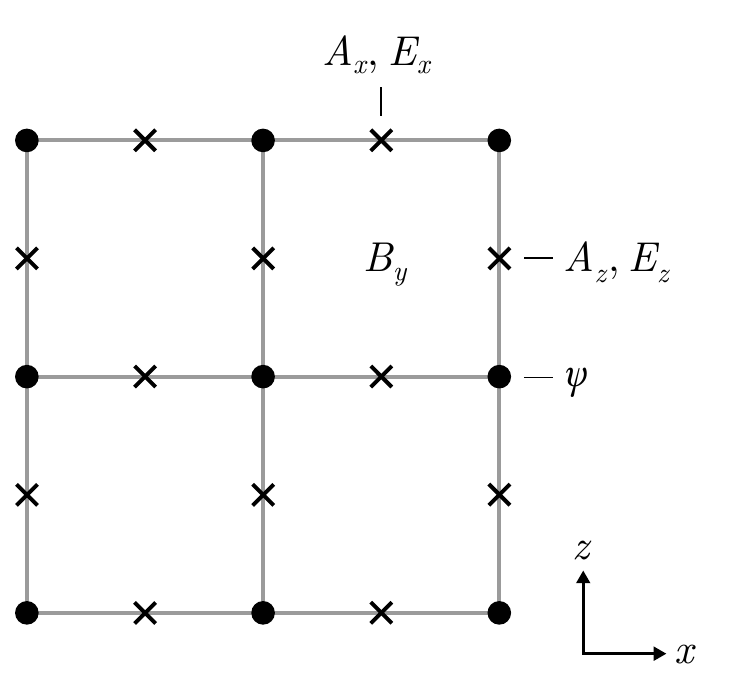}
	\caption{Discretization of the fields in the $xz$ plane.}
	\label{fig:lattice} 
\end{figure}

The discretization of the fields on the lattice is sketched in Fig.~\ref{fig:lattice}. As mentioned in the main text, the superconducting order parameter is located on the sites and the vector potential is located on the bonds. The time derivative of the vector potential leads to the electric field, which has the same spatial structure. The magnetic field inside a plaquette is defined by the discretized curl of the vector potential on the enclosing bonds. While we take the vector potential to be constant along each bond, the spatially dependent terms in the electromagnetic part of the Lagrangian describe variations of the electric and magnetic fields on length scales above the discretization length. In particular, we note that for a system at nonzero temperature, the order parameter and the electromagnetic field display spatial fluctuations, even if the driving field has a much longer wavelength. These fluctuations are captured in our simulation method.

\clearpage
\section{Comparison between the quadratic and fully nonlinear models}
\begin{figure}[!h]
	\centering
	\includegraphics[scale=1]{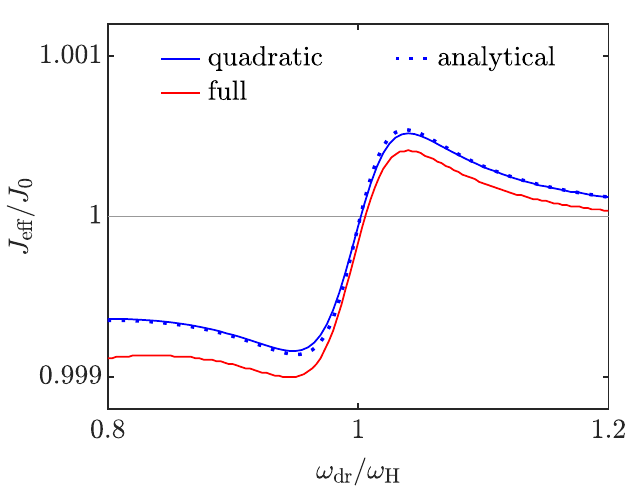}\llap{\parbox[b]{12.6cm}{\textcolor{black}{(a)}\\\rule{0ex}{4.6cm}}} \hspace{0.5cm}
	\includegraphics[scale=1]{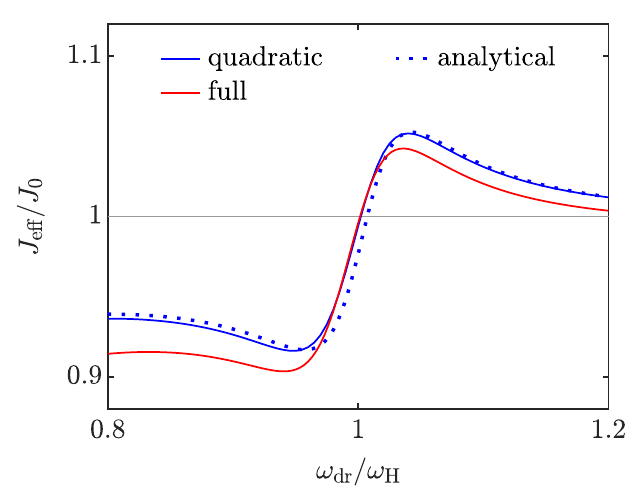}\llap{\parbox[b]{12.45cm}{\textcolor{black}{(b)}\\\rule{0ex}{4.6cm}}}
	\caption{Effective interlayer coupling for the quadratic and fully nonlinear models. The driving amplitudes are $E_0= 10~\mathrm{kV \, cm}^{-1}$ in (a) and $E_0= 100~\mathrm{kV \, cm}^{-1}$ in (b). The dotted lines indicate the analytical predictions based on the quadratic model. The model parameters are consistent with Section~\ref{sec:parameters}, with $\omega_{\mathrm{J}}/2\pi= 2~\mathrm{THz}$.}
	\label{fig:effCoupling_comp} 
\end{figure}

We refer to Eqs.~\eqref{eq:eom1_quad} and \eqref{eq:eom2_quad} as the quadratic model, which approximates the fully nonlinear model given by Eqs.~\eqref{eq:eom1_full} and \eqref{eq:eom2_full}. In Fig.~\ref{fig:effCoupling_comp}, we investigate the effect of the higher order terms on the effective interlayer coupling. We also compare the numerical results for the two models to the analytical prediction from Section~\ref{sec:analytical}. While the analytical prediction shows excellent agreement with the numerical results for the quadratic model, simulating the fully nonlinear model leads to slightly different results, even for weak driving.

\section{Depletion of the condensate}
\begin{figure}[!t]
	\centering
	\includegraphics[scale=1]{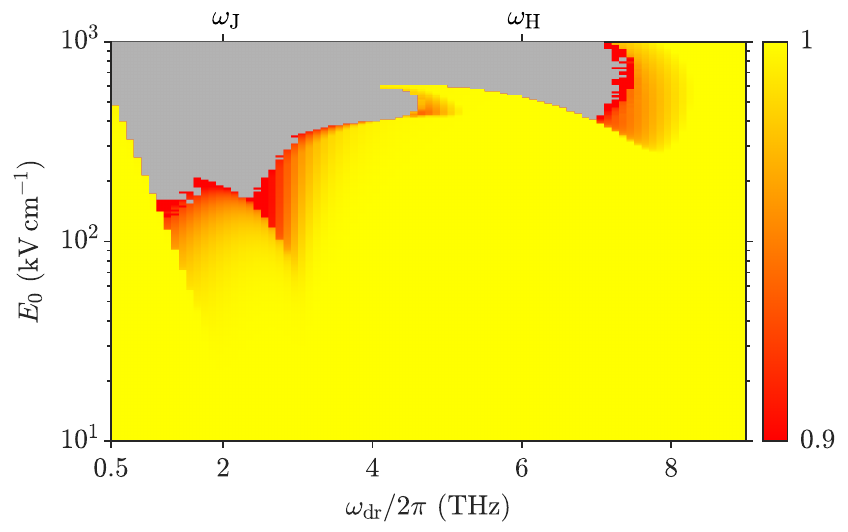}\llap{\parbox[b]{16.8cm}{\textcolor{black}{(a)}\\\rule{0ex}{4.9cm}}} \hspace{0.5cm}
	\includegraphics[scale=1]{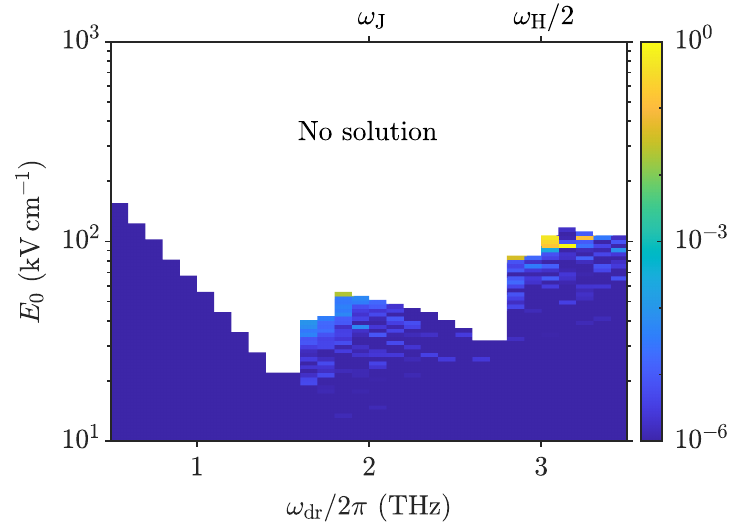}\llap{\parbox[b]{14.6cm}{\textcolor{black}{(b)}\\\rule{0ex}{4.9cm}}}
	\caption{Depletion of the condensate. (a) Dependence of the time-averaged amplitude of the order parameter $|\psi|/|\psi_0|$ on the driving frequency $\omega_{\mathrm{dr}}$ and the field strength $E_0$. (b) Absolute value of the time-averaged Higgs field $h= |\psi|/|\psi_0|-1$ as a function of driving frequency and field strength, as obtained by the harmonic balance method with ten harmonics.}
	\label{fig:depletion} 
\end{figure}

In our model, we explicitly include the drive in the time evolution of the electric field, keeping $\mu$ and $\alpha$ fixed. In fact, the gauge field effectively rescales $\mu$, which leads to a (partial) suppression of the order parameter. For simplicity, we consider the spatially homogeneous case and expand the tunneling term of the Lagrangian up to quadratic order in the vector potential:
\begin{equation}
	\mathcal{L}_{\mathrm{kin}} = - \sum_{s,\mathbf{r}} t_s | \psi_{\mathbf{r}}|^2 |1 - e^{i a_{s,\mathbf{r}}}|^2 \approx - \sum_{s,\mathbf{r}} t_s | \psi_{\mathbf{r}}|^2  a_{s,\mathbf{r}}^2 ,
\end{equation}
where $\psi_{\mathbf{r}} \equiv \psi$ and $a_{s,\mathbf{r}} \equiv a_s$ are homogeneous in space. Comparing the above expression with Eq.~(6) in the manuscript, we find an effective reduction $\mu \rightarrow \mu - \sum_s t_s a_s^2$ due to the gauge field. The effective reduction of the order parameter is displayed as a function of the driving parameters in Fig.~\ref{fig:depletion}(a). While the gauge field is small away from the resonances, it leads to a significant depletion of the condensate in the resonant regimes. Entering the heating regime results in a complete depletion of the condensate. Note that the gauge field accounts for both the external drive and the supercurrents inside the sample.

Furthermore, we scrutinize the dynamical stability of the Higgs field and the vector potential using the harmonic balance method. Given a set of nonlinear ordinary differential equations with periodically changing parameters, the harmonic balance method maps the problem into an algebraic one by expanding solutions by multiple harmonics. The obtained algebraic equation is solved by the Newton or secant method. In order to solve the fully nonlinear equations of motion \eqref{eq:eom1_full} and \eqref{eq:eom2_full}, we use the Krylov-Newton method with ten harmonics in Mousai \cite{Slater2017}.

In Fig.~\ref{fig:depletion}(b), we plot the absolute value of the time-averaged Higgs field for various driving frequencies $\omega_\text{dr}$ and amplitudes $E_0$ (corresponding to Fig.~3 in the main text). Due to the nonlinearity of the equation of motion, the system may show multistability; the obtained solution depends on the initial condition. Here, at each frequency, we sweep from weak to strong driving using the preceding calculation as the initial condition for the next calculation. We confirm the stability of the obtained solutions by adding slight noise to the initial conditions.

The blank area corresponds to the case where the Krylov-Newton method fails to obtain periodic steady solutions. The strong instability appears around $\omega_\text{dr} \simeq \omega_\text{J}$ and $\omega_\text{H}/2$, which agrees with the numerical solutions in the main text. The harmonic balance method overestimates the instability regimes due the difficulty of solving nonlinear algebraic equations accurately. At higher frequencies near $\omega_\text{H}$, we find that the multistability is more prominent, and that the instability needs large driving amplitude. This regime corresponds to the heating regime identified via spectral entropy or depletion. The deviation between the harmonic balance method and the numerical solutions may come from the fact that chaotic solutions cannot be obtained by the harmonic balance method due to its harmonic ansatz.

\section{The role of the damping coefficients and the capacitive coupling constant}
\begin{figure}[!h]
	\centering
	\includegraphics[scale=1]{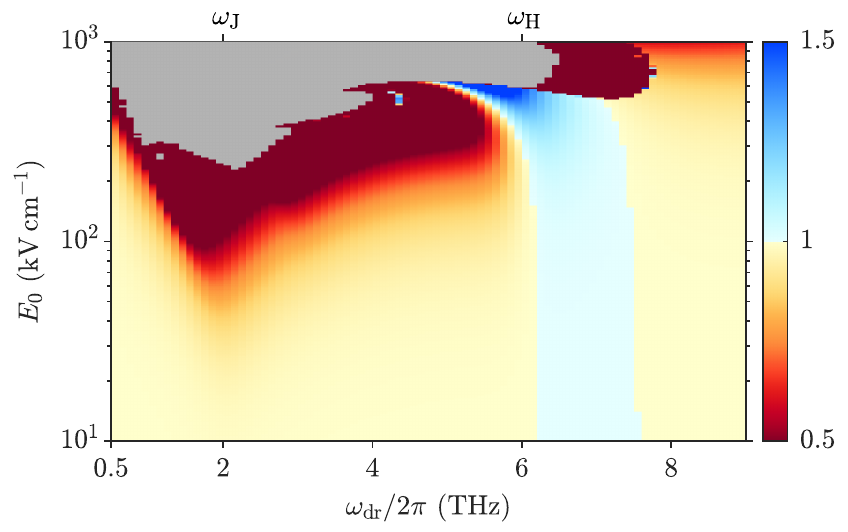}
	\caption{Dependence of the effective interlayer coupling $J_{\mathrm{eff}}/J_0$ on the driving frequency $\omega_{\mathrm{dr}}$ and the field strength $E_0$. The gray area marks the heating regime. The damping coefficients are $\gamma_{\mathrm{H}}/2\pi= \gamma_{\mathrm{J}}/2\pi= 1~\mathrm{THz}$.}
	\label{fig:condScan} 
\end{figure}

In this section, we investigate how the Higgs mode mediated enhancement of interlayer transport depends on the damping coefficients and the capacitive coupling constant. In analogy to Fig.~3 in the main text, we numerically evaluate the renormalized interlayer coupling for different driving parameters, but we choose larger damping coefficients $\gamma_{\mathrm{H}}/2\pi= \gamma_{\mathrm{J}}/2\pi= 1~\mathrm{THz}$. We see in Fig.~\ref{fig:condScan} that the enhancement and reduction regimes are similar as before. The magnitude of the enhancement is generally smaller, but it is still possible to enhance the effective coupling by $50\%$ when using sufficently strong fields.

\begin{figure}[!h]
	\centering
	\includegraphics[scale=1]{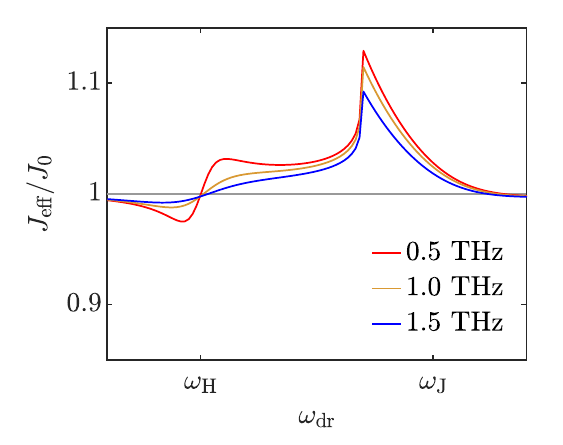}\llap{\parbox[b]{10.8cm}{\textcolor{black}{(a)}\\\rule{0ex}{4cm}}}
	\includegraphics[scale=1]{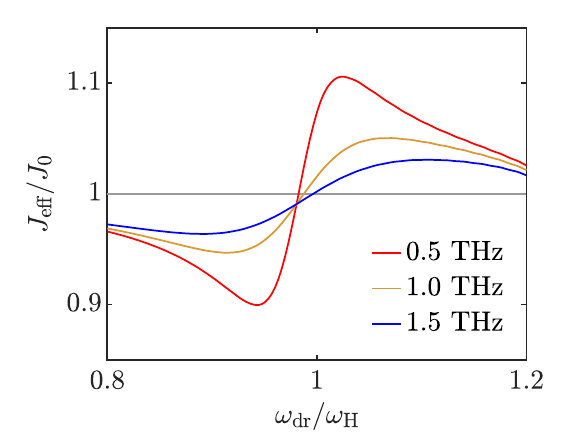}\llap{\parbox[b]{10.8cm}{\textcolor{black}{(b)}\\\rule{0ex}{4cm}}}
	\caption{Effective interlayer coupling for various values of $\gamma_{\mathrm{H}}/2\pi$. The Higgs frequency $\omega_{\mathrm{H}}/2\pi= 6~\mathrm{THz}$ and $\gamma_{\mathrm{J}}/2\pi= 0.5~\mathrm{THz}$ are kept fixed. The choice of the driving amplitude depends on the plasma frequency: (a) $E_0= 20~\mathrm{kV \, cm}^{-1}$ for $\omega_{\mathrm{J}}/2\pi= 9~\mathrm{THz}$ and (b) $E_0= 100~\mathrm{kV \, cm}^{-1}$ for $\omega_{\mathrm{J}}/2\pi= 15~\mathrm{THz}$.}
	\label{fig:varGammaSc} 
\end{figure}

\begin{figure}[!h]
	\centering
	\includegraphics[scale=1]{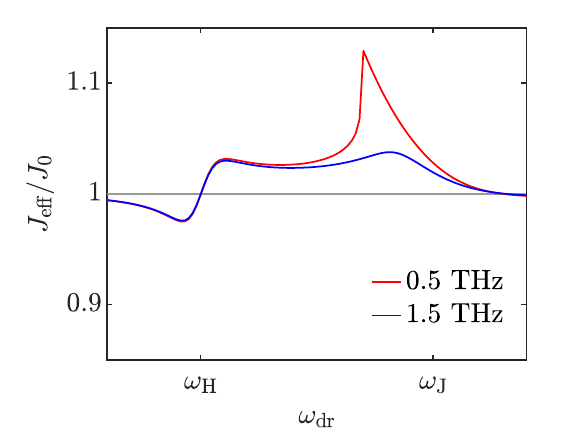}\llap{\parbox[b]{10.8cm}{\textcolor{black}{(a)}\\\rule{0ex}{4cm}}}
	\includegraphics[scale=1]{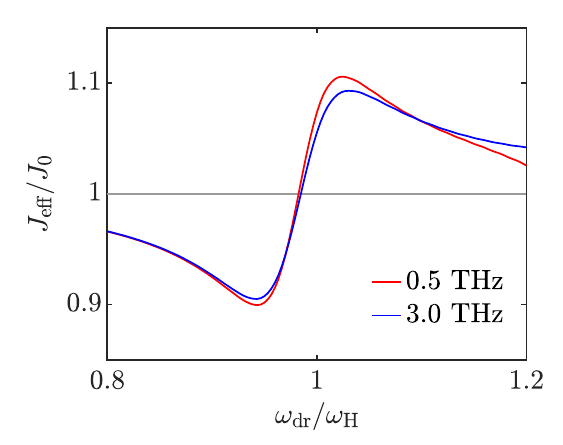}\llap{\parbox[b]{10.8cm}{\textcolor{black}{(b)}\\\rule{0ex}{4cm}}}
	\caption{Effective interlayer coupling for various values of $\gamma_{\mathrm{J}}/2\pi$. The Higgs frequency $\omega_{\mathrm{H}}/2\pi= 6~\mathrm{THz}$ and $\gamma_{\mathrm{H}}/2\pi= 0.5~\mathrm{THz}$ are kept fixed. The choice of the driving amplitude depends on the plasma frequency: (a) $E_0= 20~\mathrm{kV \, cm}^{-1}$ for $\omega_{\mathrm{J}}/2\pi= 9~\mathrm{THz}$ and (b) $E_0= 100~\mathrm{kV \, cm}^{-1}$ for $\omega_{\mathrm{J}}/2\pi= 15~\mathrm{THz}$.}
	\label{fig:varGammaEl} 
\end{figure}

To further scrutinize the role of the damping coefficients, we present the effective interlayer coupling for different values of $\gamma_{\mathrm{H}}$ and $\gamma_{\mathrm{J}}$ for the the two cases of the ratio $\omega_{\mathrm{J}}/\omega_{\mathrm{H}}$ that are not shown in the main text. Figure~\ref{fig:varGammaSc} displays the effect of increasing the damping coefficient of the Higgs mode. In both cases, the renormalization of the effective interlayer coupling is visible for higher values of $\gamma_{\mathrm{H}}$, but the effect is notably weaker. Higher values of $\gamma_{\mathrm{J}}$, on the other hand, do not significantly reduce the enhancement of the effective interlayer coupling for driving frequencies near the Higgs frequency, as shown in Fig.~\ref{fig:varGammaEl}. Interestingly, the enhancement near the plasma frequency for $\omega_{\mathrm{H}} < \omega_{\mathrm{J}} < \sqrt{3} \omega_{\mathrm{H}}$ is more severely affected by $\gamma_{\mathrm{J}}$ than by $\gamma_{\mathrm{H}}$.

\begin{figure}[!h]
	\centering
	\includegraphics[scale=1]{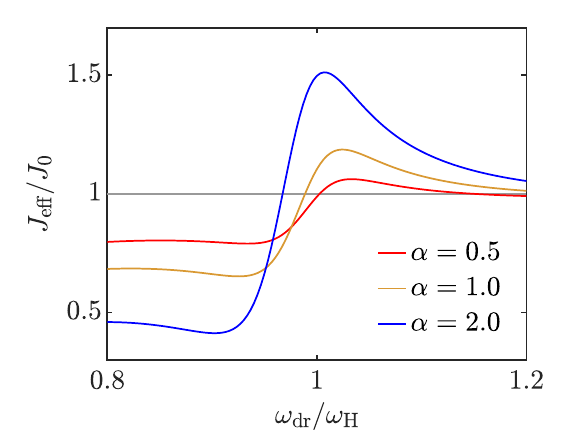}\llap{\parbox[b]{10.8cm}{\textcolor{black}{(a)}\\\rule{0ex}{4cm}}}
	\includegraphics[scale=1]{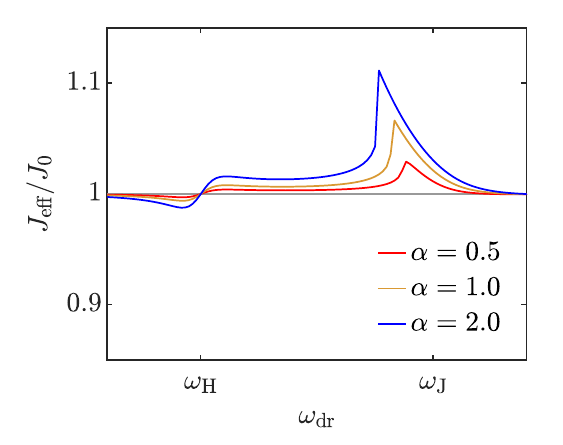}\llap{\parbox[b]{10.8cm}{\textcolor{black}{(b)}\\\rule{0ex}{4cm}}}
	\includegraphics[scale=1]{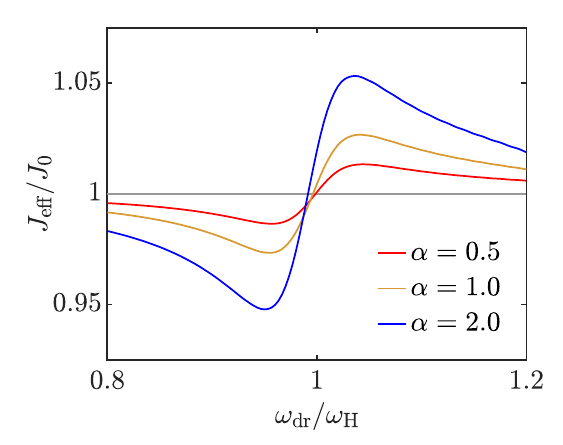}\llap{\parbox[b]{10.8cm}{\textcolor{black}{(c)}\\\rule{0ex}{4cm}}}
	\caption{Effective interlayer coupling for various values of the capacitive couling constant $\alpha$. The Higgs frequency is $\omega_{\mathrm{H}}/2\pi= 6~\mathrm{THz}$. The choice of the driving amplitude depends on the plasma frequency: (a) $E_0= 200~\mathrm{kV \, cm}^{-1}$ for $\omega_{\mathrm{J}}/2\pi= 2~\mathrm{THz}$, (b) $E_0= 10~\mathrm{kV \, cm}^{-1}$ for $\omega_{\mathrm{J}}/2\pi= 9~\mathrm{THz}$, and (c) $E_0= 50~\mathrm{kV \, cm}^{-1}$ for $\omega_{\mathrm{J}}/2\pi= 15~\mathrm{THz}$.}
	\label{fig:varAlpha} 
\end{figure}

\pagebreak
To modify the capacitive coupling constant $\alpha$, we vary the interaction strength $g$ and adjust the tunneling coefficient $t_z$ such that the previous values of the plasma frequency $\omega_{\mathrm{J}}$ are recovered. It is evidenced by Fig.~\ref{fig:varAlpha} that the Higgs mode mediated enhancement is qualitatively not affected by the value of $\alpha$, which is of the order of 1 in the cuprates \cite{Machida2004}. As expected, the enhancement is most pronounced for the largest $\alpha$.

\clearpage
\section{Higgs mode mediated enhancement at nonzero temperature}
Here, we consider a monolayer cuprate superconductor at nonzero temperature. For this purpose, we simulate a three-dimensional system of $48 \times 48 \times 4$ sites with the parameters specified in the Appendix of the main text, choosing $t_z= 9.44 \times 10^{-4}~\mathrm{meV}$. The equations of motion read
\begin{align}
\partial_t^2 \psi_{\mathbf{r}} &= \frac{1}{K \hbar^2} \frac{\partial \mathcal{L}}{\partial \psi_{\mathbf{r}}^*} - \gamma_{\mathrm{H}} \partial_t \psi_{\mathbf{r}} + \xi_{\mathbf{r}} , \\
\partial_t^2 A_{s, \mathbf{r}} &= \frac{1}{\epsilon_s \epsilon_0} \frac{\partial \mathcal{L}}{\partial A_{s, \mathbf{r}}} - \gamma_{\mathrm{J}} \partial_t A_{s, \mathbf{r}} + \eta_{s, \mathbf{r}} ,
\end{align}
where $\xi_{\mathbf{r}}$ and $\eta_{s, \mathbf{r}}$ represent the thermal fluctuations of the superconducting order parameter and the vector potential, respectively. These Langevin noise terms have a white Gaussian distribution with zero mean. To satisfy the fluctuation-dissipation theorem, we take the noise of the order parameter as
\begin{align}
\langle \mathrm{Re} \{{\xi_{\mathbf{r}} (t)}\} \mathrm{Re} \{{\xi_{\mathbf{r'}} (t')}\} \rangle &= \frac{\gamma_{\mathrm{H}} k_{\mathrm{B}} T}{K \hbar^2 V_0} \delta_{\mathbf{r}\mathbf{r'}} \delta(t-t') \, , \\
\langle \mathrm{Im} \{{\xi_{\mathbf{r}} (t)}\} \mathrm{Im} \{{\xi_{\mathbf{r'}} (t')}\} \rangle &= \frac{\gamma_{\mathrm{H}} k_{\mathrm{B}} T}{K \hbar^2 V_0} \delta_{\mathbf{r}\mathbf{r'}} \delta(t-t') \, , \\
\langle \mathrm{Re} \{{\xi_{\mathbf{r}} (t)}\} \mathrm{Im} \{{\xi_{\mathbf{r'}} (t')}\} \rangle &= 0 .
\end{align}
The noise correlations for the vector potential are
\begin{align}
\langle \eta_{x,\mathbf{r}} (t) \eta_{x,\mathbf{r'}} (t') \rangle &= \frac{2 \gamma_{\mathrm{J}} k_{\mathrm{B}} T}{\epsilon_x \epsilon_0 V_0} \delta_{\mathbf{r}\mathbf{r'}} \delta(t-t') , \\
\langle \eta_{y,\mathbf{r}} (t) \eta_{y,\mathbf{r'}} (t') \rangle &= \frac{2 \gamma_{\mathrm{J}} k_{\mathrm{B}} T}{\epsilon_y \epsilon_0 V_0} \delta_{\mathbf{r}\mathbf{r'}} \delta(t-t') , \\
\langle \eta_{z,\mathbf{r}} (t) \eta_{z,\mathbf{r'}} (t') \rangle &= \frac{2 \gamma_{\mathrm{J}} k_{\mathrm{B}} T}{\epsilon_z \epsilon_0 V_0} \delta_{\mathbf{r}\mathbf{r'}} \delta(t-t') .
\end{align}

\begin{figure}[!h]
	\centering
	\includegraphics[scale=1]{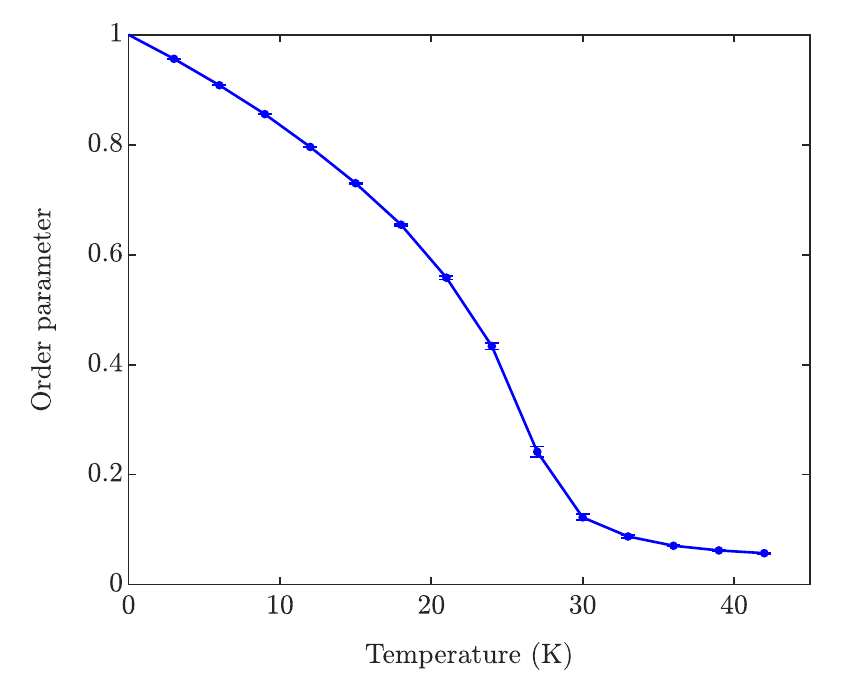}\llap{\parbox[b]{16.6cm}{\textcolor{black}{(a)}\\\rule{0ex}{6.5cm}}}
	\includegraphics[scale=1]{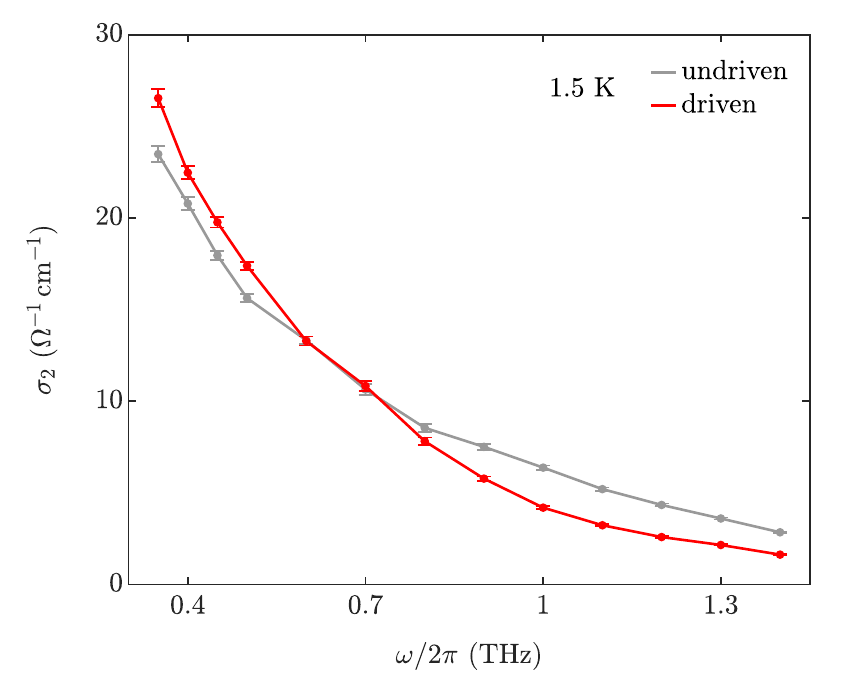}\llap{\parbox[b]{16.6cm}{\textcolor{black}{(b)}\\\rule{0ex}{6.5cm}}}
	\caption{Finite-temperature results. (a) Phase transition of a system of $48 \times 48 \times 4$ sites. (b) Higgs mode mediated enhancement of the imaginary $c$-axis conductivity for low frequencies at $T= 1.5~\mathrm{K}$. The driving frequency is $\omega_{\mathrm{dr}} \approx 1.1 \, \omega_{\mathrm{H}}$. In both panels, the error bars indicate the standard errors of the ensemble averages.}
	\label{fig:finiteTemp} 
\end{figure}

The thermal equilibrium at a given temperature is established as follows. We initialize the system in its ground state at $T=0$ and let the dynamics evolve without external driving, influenced only by thermal fluctuations and dissipation. To characterize the phase transition, we introduce the order parameter
\begin{equation}
	O= \frac{ \Big|\sum_{l,m,n} \psi_{l,m,n+1}^* \psi_{l,m,n} \, \mathrm{e}^{\mathrm{i} a_{l,m,n}^z} \Big| }{\sum_{l,m,n} |\psi_{l,m,n}|^2}.
\end{equation}
The order parameter measures the gauge-invariant phase coherence of the condensate across different layers. In our simulations, this quantity converges to a constant after 10 ps of free time evolution, indicating that thermal equilibrium is reached. For each trajectory, the order parameter is evaluated from the average of 200 measurements within a time interval of 2 ps. Finally, we take the ensemble average of 100 trajectories. As shown in Fig.~\ref{fig:finiteTemp}(a), the temperature dependence of the order parameter is reminiscent of a second order phase transition. Due to the finite size of the simulated system, the order parameter converges to a plateau with nonzero value for high temperatures. Instead of a sharp discontinuity, one finds a distinct crossover at $T_c \sim 30~{\mathrm{K}}$.

To obtain the $c$-axis conductivity at nonzero temperature, we add a probe term to the equations of motion for $A_{z,\mathbf{r}}$. Then, we compute $\sigma(\omega)$ as the ratio of the sample averages of $J_z(\omega)$ and $E_z(\omega)$ before taking the ensemble average of several hundred trajectories.
In Fig.~\ref{fig:finiteTemp}(b), we present an example of Higgs mode mediated enhancement of interlayer transport at $T= 1.5~\mathrm{K}$. Applying an optical drive with the frequency $\omega_{\mathrm{dr}}/2\pi= 7.4~\mathrm{THz}$ and the field strength $E_0= 400~\mathrm{kV \, cm^{-1}}$, we find a low-frequency enhancement of $\sigma_2$ by $\sim 10\%$. Note that the $c$-axis plasma frequency and the Higgs frequency are $\omega_{\mathrm{J}}/2\pi \approx 2~\mathrm{THz}$ and $\omega_{\mathrm{H}}/2\pi \approx 6.7~\mathrm{THz}$, respectively, at this temperature.

\bibliography{biblio}